\documentclass[aps,prb,twocolumn,groupedaddress]{revtex4}
\usepackage[dvips]{graphics}
\usepackage{vmargin}
\usepackage[dvips]{color}
\usepackage{amsmath}
\usepackage{subfigure}

\setpapersize{USletter}
\setmarginsrb{1in}{1.5in}{1in}{0.5in}{0in}{0.2in}{0in}{0in}

\begin{document}

\title{First-principle studies of the spin-orbit and
the Dzyaloshinskii-Moriya interactions in the \{Cu$_3$\} single-molecule magnet}

\author{J.F. Nossa}
\author{M.F. Islam}
\author{C.M. Canali}
\affiliation{School of Computer Science, Physics and Mathematics, Linnaeus University, Kalmar-Sweden}
\author{M.R. Pederson}
\affiliation{Naval Research Laboratory, Washington DC, USA}
\date{\today}

\begin{abstract}
Frustrated triangular molecule magnets such as \{Cu$_3$\} are
characterized by two degenerate S=1/2 ground-states with opposite
chirality. Recently it has been proposed theoretically [PRL {\bf
101}, 217201 (2008)] and verified by {\it ab-initio} calculations
[PRB {\bf 82}, 155446 (2010)] that an external electric field can
efficiently couple these two chiral spin states, even in the absence
of spin-orbit interaction (SOI). The SOI is nevertheless important,
since it introduces a splitting in the ground-state manifold via the
Dzyaloshinskii-Moriya interaction. In this paper we present a
theoretical study of the effect of the SOI on the chiral states
within spin density functional theory. We employ a
recently-introduced Hubbard model approach to elucidate the
connection between the SOI and the Dzyaloshinskii-Moriya
interaction. This allows us to express the  Dzyaloshinskii-Moriya
interaction constant $D$ in terms of the microscopic Hubbard model
parameters, which we calculate from first-principles. The small
splitting that we find for the \{Cu$_3$\} chiral state energies
($\Delta \approx 0.02$ meV) is consistent with experimental results.
The Hubbard model approach adopted here also yields a better
estimate of the isotropic exchange constant than the ones obtained
by comparing total energies of different spin configurations. The method used here for calculating 
the DM interaction unmasks its simple fundamental origin which is the off-diagonal spin-orbit interaction
between the generally multireference vacuum state and single-electron excitations out of those states
\end{abstract}


\maketitle

%%%%%%%%%%%%%%%%%%%%%%%%%%%%%%%%%%%%%%%%%%%%%%%%%%%%%%%%%%%%%
%                                                           %
%       section               Introduction
%                                                           %
%%%%%%%%%%%%%%%%%%%%%%%%%%%%%%%%%%%%%%%%%%%%%%%%%%%%%%%%%%%%%

\section{Introduction}
\label{intro}

In the last twenty years single-molecule magnets (SMMs) have been
widely studied both for their fundamental physical
properties~\cite{Gatteschi2006}, and for possible appli\-cations in
magnetic storage and quantum
information.\cite{Leuenberger2001,Lehmann2007} Unlike traditional
bulk magnetic mate\-rials, molecular magnetic  materials can be
magnetized in a magnetic field without any interaction between the
individual molecules. This magnetization is a property of the
molecules themselves. The magnetization occurs because of the large
ground-state spin and the large easy-axis magnetic anisotropy
barrier separating spin-up and the spin-down states. In principle it
is possible to store and manipulate information in one SMM.
Furthermore the two quantum states representing the two possible
spin orientations can be used to build a quantum qubit. Whether used
as classical magnetic storage units or as quantum coherent elements,
the crucial requirement in both cases is the ability to control and
manipulate the magnetic states of the SMM in an efficient way.
Manipulation by magnetic fields is straightforward but, in practice,
cannot be realized with molecular-size spatial resolution and at
fast temporal scales. Unlike magnetic fields, electric fields are
easy to produce, quickly switched and can be applied locally at the
nano and molecular scale. Therefore manipulation of the properties
of SMMs by external electric fields is an attractive and promising
alternative.\cite{Trif2008}

Although electric fields do not directly couple to spins, electric
manipulation of the spin states is possible indirectly via
spin-orbit coupling. This requires the presence of a strong
spin-orbit coupling such that the electric field can effectively
flip the spin states by acting on the the orbital part of the
spin-orbitals. When SMMs are involved, this is not the most
efficient mechanism, since the relative strength of spin-orbit
interaction scales like the volume of the molecule.

Recently, a different mechanism of spin-electric coupling in
antiferromagnetic SMMs, characterized by lack of inversion symmetry
and spin frustration, has been proposed.\cite{Trif2008} The best
example of such a system is a triangular spin $s = 1/2$ ring with
antiferromagnetic coupling, realized for example in the \{Cu$_3$\}
SMM. The low energy physics of this system can be described by a
three-site spin $s= 1/2$ Heisenberg Hamiltonian whose ground-state
manifold is composed of two degenerate (total) spin $S=1/2 $
doublets, with wave functions represented by
\begin{equation}
|\chi_{\pm}, S_z= +\frac{1}{2}\rangle =\frac{1}{\sqrt{3}} \big (|\downarrow
\uparrow \uparrow\rangle + \epsilon_{\pm}
|\uparrow\downarrow\uparrow\rangle + \epsilon_{\mp}
|\uparrow\uparrow\downarrow\rangle\big ) \label{eq:chi_pm_sz+}\;,
\end{equation}
\begin{equation}
|\chi_ {\pm}, S_z= -\frac{1}{2}\rangle =\frac{1}{\sqrt{3}} \big
(|\uparrow\downarrow \downarrow \rangle + \epsilon_{\pm}
|\downarrow\uparrow\downarrow\rangle + \epsilon_{\mp}
|\downarrow\downarrow\uparrow\rangle\big ) \label{eq:chi_pm_sz-}\;,
\end{equation}
where the many-body states $|\sigma_1\sigma_2\sigma_3\rangle$
are products of spin-orbital states  $\sigma_i =(\uparrow,
\downarrow), i = 1,2,3$ localized on the three magnetic ions of the
molecules, and $\epsilon_{\pm}= \exp\left( \pm 2\pi i / 3\right)$.
The four states $|\chi_{\pm}, S_z = \pm 1/2 \rangle$ in
Eqs.~(\ref{eq:chi_pm_sz+}), (\ref{eq:chi_pm_sz-}) are labeled by the
eigenvalues  $S_z=\pm 1/2$ of the $z$-component of the total spin,
and by the chirality quantum number $\chi_{\pm} = \pm 1$, that is,
the eigenvalues of the chiral operator
\begin{equation}
C_z= \frac{4}{\sqrt{3}} {\mathbf s}_1\cdot {\mathbf s}_2 \times
{\mathbf s}_3 \label{eq:chiraloperator}\;.
\end{equation}

An electric field couples to the SMM through $e {\bf E}\cdot {\bf
R}$, where $e$ is the electron charge and ${\bf R} = \sum_{i= 1}^3
{\bf r}_i$. The two spin-orbital states $|\chi_{\pm}, S_z\rangle$,
cha\-racterized by opposite chirality and equal spin projection, form
the basis of a two-dimensional $E'$ irreducible representation of
$D_{3h}$. General group theory arguments then guarantee that the
matrix elements, $e\langle \chi_{+1}, S_z| X_-|\chi_{-1}, S_z\rangle =
e\langle \chi_{-1}, S_z| X_+|\chi_{+1}, S_z\rangle = 2id\neq 0$, where
$X_{\pm}\equiv \pm X + iY$ are the in-plane components of $\bf R$,
which also transform as the two-dimensional irreducible representation $E'$. Here $d$ is a
real number that is refereed to as spin electric-dipole coupling.
It follows that, due to these non-zero matrix ele\-ments, an electric
field can cause transitions bet\-ween two ground state wavefunctions of opposite
chirality but with same $S_z$.

\begin{figure}[h]
{\resizebox{3in}{1.7in}{\includegraphics{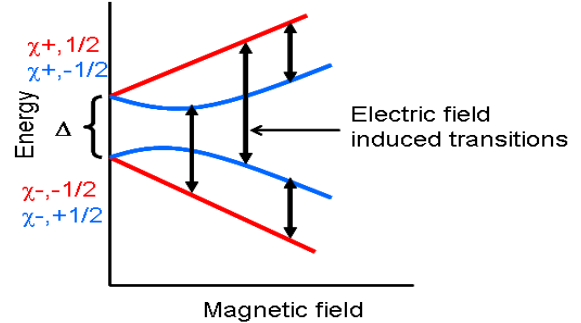}}}
\caption{ (Color online) Schematic diagram of electric-field-induced transitions between states of
different chirality belonging to the spin $S= 1/2$ ground-state manifold of a triangular
antiferromagnet.
$\Delta$ represents zero-field splitting of the
chiral states due to Dzyaloshinskii-Moriya interaction.}
\label{spin-elec}
\end{figure}

The observation of such electric-field induced transitions from one
chiral state to another requires that the degeneracies between these
states be lifted. The anisotropic Dzyaloshinskii-Moriya (DM)
interaction plays a crucial role in that, it provides one possible
mechanism that lifts the degeneracies between states of different
chirality without mixing them, as shown in Fig.~\ref{spin-elec}.
More in general, the presence of DM interaction provides a
mechanism to control the size of quantum entanglement in magnetic
trimers as a function of the temperature and external magnetic
field\cite{florez_JMMM_2012}. Experimentally the DM-induced
splitting in \{Cu$_3$\} is estimated to be small (approximately 0.5
K\cite{choi06}).

Recently\cite{islam2010} we have investigated the details of the
electronic properties of the \{Cu$_3$\} SMM, which is one of the
most promising triangular spin 1/2 molecules where the spin-electric
effect can be rea\-lized. In particular, we introduced a scheme to
evaluate the strength of spin electric-dipole coupling $d$ using
{\it ab-initio} methods. However, the value of the anisotropic
DM-exchange constant interaction, which is responsible of the GS
zero-field splitting, has not yet been calculated. The purpose of
this work is to calculate this splitting by {\it ab-initio} me\-thods.
In order to achieve this goal, we analyze the microscopic origin of
the DM interaction via a Hubbard model approach in the presence of
spin-orbit integration, which is the correct minimal model to
describe both spin and charge fluctuations of these strongly
correlated electron systems. At half-filling and in the large
Hubbard $U$ limit, spin-dependent virtual ho\-pping processes, induced
by the spin-orbit interaction, give rise to an anisotropic exchange
interaction.\cite{Trif2010} There is a close analogy with the
isotropic Heisenberg exchange interaction obtained in second-order
perturbation theo\-ry in the spin-independent ho\-pping perturbation.
Beside elucidating the physical mechanism leading to the anisotropic
DM exchange interaction, this approach provides a very convenient
prescription on how to extract the DM exchange cons\-tant from
first-principle calculations, which we have ca\-rried out for
\{Cu$_3$\}.

This paper is organized as follows. In
Sec.~\ref{sec:GeneralpropertiesoftheDMinteraction} we discuss the
general properties of the DM interaction. The Hubbard model approach
for calculating DM vector, adopted in this work, is discussed in
Sec.~\ref{methods:3}. In
Sec.~\ref{sec:AbinitiocalculationoftheDMvector} we discuss details
of extracting Hubbard model parameters from our {\it ab-initio}
calculations. In Sec.~\ref{sec:othermethods} we discuss other
methods that are usually employed for calculating the DM vector.
Finally in Sec.~\ref{summary} we present a summary of our work.

%%%%%%%%%%%%%%%%%%%%%%%%%%%%%%%%%%%%%%%%%%%%%%%%%%%%%%%%%%%%%%%%%%%%%%%%%%%%%%%%%
%                                                                               %
%  section  The Dzyaloshinskii-Moriya interaction in frustrated antiferromagnetic spin rings
%                                                                               %
%%%%%%%%%%%%%%%%%%%%%%%%%%%%%%%%%%%%%%%%%%%%%%%%%%%%%%%%%%%%%%%%%%%%%%%%%%%%%%%%%
\section{The Dzyaloshinskii-Moriya interaction in frustrated antiferromagnetic spin rings}
\label{DM_theory}

%%%%%%%%%%%%%%%%%%%%%%%%%%%%%%%%%%%%%%%%%%%%%%%%%%%%%%%%%%%%%%%%%%%%%%%%%%%%%%%%%
%                                                                               %
%  subsection         General properties of the DM interaction
%                                                                               %
%%%%%%%%%%%%%%%%%%%%%%%%%%%%%%%%%%%%%%%%%%%%%%%%%%%%%%%%%%%%%%%%%%%%%%%%%%%%%%%%%
\subsection{General properties of the DM interaction}
\label{sec:GeneralpropertiesoftheDMinteraction}

The Dzyaloshinskii-Moriya (DM) interaction is an anisotropic
exchange interaction resulting from the interplay of the Coulomb
interaction and the spin-orbit coupling in systems of low crystal
symmetry. The DM interaction is an important effect for many
magnetic systems and plays a crucial role in determining the
zero-field splitting of energy levels. An anisotropic exchange
interaction of the form
\begin{equation}
{\bf D}_{12}\cdot {\bf S}_1 \times {\bf S}_2\;,
\label{dmi}
\end{equation}
which is linear in the spin-orbit interaction, was first put forward
by Dzyaloshinskii on the basis of symme\-try
considerations.\cite{Dzyaloshinskii_1958}  Later
Moriya\cite{Moriya_prl_1960, Moriya1960} provided a mechanism for
this interaction by exten\-ding Anderson's theory of
superexchange\cite{PW_Anderson_1959} to include the effect of
spin-orbit coupling. Let us consider for simplicity two ``magnetic
ions", $\bf R$ and $\bf R'$, each occupied by a single electron in
the ground state. Second-order perturbation theory in the ho\-pping
Hamiltonian $H_t$ coupling the two sites gives rise to an isotropic
antiferromagnetic interaction with exchange constant $J = 2 t_{\bf R
R'}^2/U$, where $t_{\bf R R'}$ is a spin-independent ho\-pping
integral and $U$ is the energy required to transfer an electron from
$\bf R$ to $\bf R'$. When spin-orbit interaction $H_{\rm SOI}$ is
included, similar second-order processes can gene\-rate an anisotropic
exchange interaction in the form of Eq.~(\ref{dmi}), with $D  \sim
t_{\bf R R'} b_{\bf R R'}/U$ where $b_{\bf R R'}$ is a SOI-induced
or SOI-dependent ho\-pping integral. To lowest-order, $b_{\bf R R'}$
is just the matrix elements of the $H_{\rm SOI}$ between two
orbitals localized at $\bf R$ and $\bf R'$. This is the dominant
contribution to $D$. In case that at each site more than one orbital
$|{\bf R}, \mu\rangle\;,  \mu = 1, 2, \cdots $ plays a role,
higher-order terms such as $b_{\bf R R'} = t_{\bf R R'} \langle {\bf
R}, \mu |H_{\rm SOI}|{\bf R}', \mu' \rangle/\Delta E_{\mu, \mu'}$ are
possible, making the corresponding $D$ effectively a third-order
coupling in the perturbations $H_t$ and $H_{\rm SOI}$. It turns out
that $D \simeq (\Delta g/g)$, where $g$ is the free-electron
gyromagnetic ratio and $\Delta g$ the deviation from $g$ induced by
SOI.\cite{Moriya1960}

As shown by Moriya, other terms linear in the SOI contribute to the
anisotropic exchange of the form of Eq.~(\ref{dmi}). The second most
important contribution is also a second-order term resulting from
SOI and {\it direct} inter-atomic exchange interaction $J^{\rm
ex}({\bf R, R'})$. In antiferromagnetic crystals this term is
$J^{\rm ex}({\bf R, R'})/J$ times smaller than the second-order
contribution proportional to $t_{\bf R R'} b_{\bf R R'}$. Finally,
third-order contributions  to $D$ include the ho\-pping terms twice
and the intra-atomic exchange constant $J_0$. They are $J_0/U$
smaller than second-order terms.

The DM exchange vector $ \bf D$ vanishes when the symmetry of the
crystal is high. This is the case, for example, when the point
located halfway between the two magnetic ions in a unit cell is a
center of inversion. In low-dimensional crystals where $D \neq 0$,
the anisotropic exchange is typically the most important anisotropic
contribution between spins. The DM interaction favors non-collinear
spin configurations, with typical canted spins. As such, it
determines the spin arrangements and it is responsible for the weak
ferromagnetism observed in some predo\-minantly antiferromagnetic
crystals such as $\alpha$-Fe$_2$O$_2$. The tendency toward canted
spin configurations is most-easily seen by minimizing the energy in
Eq.~(\ref{dmi}) for two classical spins, when the DM vector $\bf D$
is, for example, perpendicular to the line joining the two ions. It
can be shown that the minimum ener\-gy corresponds to a spin
configuration where both spin are perpendicular to each other and to
the direction of $\bf D$. Similar conclusions can be obtained by
analyzing the same system quantum mechanically. The DM interactions
is also responsible for proposed non-collinear spin configurations
in magnetic clusters engineered by STM techniques on insulating
surfaces.\cite{cyrus07,lichtenstein09}

%%%%%%%%%%%%%%%%%%%%%%%%%%%%%%%%%%%%%%%%%%%%%%%%%%%%%%%%%%%%%
%                                                           %
%       subsection  The DM interaction for antiferromagnetic spin rings within a Hubbard model approach
%                                                           %
%%%%%%%%%%%%%%%%%%%%%%%%%%%%%%%%%%%%%%%%%%%%%%%%%%%%%%%%%%%%%
\subsection{The DM interaction for antiferromagnetic spin rings within a Hubbard model approach}
\label{methods:3}

In this section we specialize the previous discussion to the
case of an antiferromagnetic spin triangle,
and show how the DM interaction can be derived microscopically from
a Hubbard model at half filling, in the presence of spin-orbit
interaction.

As mentioned in the introduction, the low-energy magnetic properties
of \{Cu$_3$\} are well-described by an isotropic antiferromagnetic
Heisenberg model
\begin{equation}
%H_{\rm H} = -2 \sum_{\left\langle i,j \right\rangle } J_{ij}{\bf
H_{\rm H} = \sum_{\left\langle i,j \right\rangle } J_{ij}{\bf
s}_i \cdot {\bf s}_j\;, \ \ \ J_{ij}>0\;, \label{eq:afmh}
\end{equation}
where ${\bf s}_i$ are spin vector operators of magnitude $s_i =1/2$,
predominately localized at the three Cu sites. If the small
distortion from a perfect equilateral arrangement of the three Cu
atoms is neglected, the three exchange constants are the same,
$J_{ij} = J$. DFT calculations\cite{islam2010} find $J\approx $ 3.7
meV. The GS manifold comprises two spin $S=1/2$ doublets, which can
be represented by the two chiral states given in
Eqs.~(\ref{eq:chi_pm_sz+}),~(\ref{eq:chi_pm_sz-}), or any two
orthogonal linear combination of these. The spin $S=3/2$
excited-state multiplet is separated by the GS by an energy of order
$J$.

It is well-known that the AFM Heisenberg model represents an
effective low-energy spin model that can be derived from an
underlying Hubbard model at half-filling in the large $t/U$ limit.
The choice of the best minimal model capturing the essential
microscopic features of the electronic system is often a complex
task, particularly when the exchange interaction between the
magnetic ions is mediated via several paths involving non-magnetic
ions, as for the case of \{Cu$_3$\}. We will neglect these
complications  and assume that an effective one-band Hubbard model
suffices for this purpose. We will see that our first-principles
calculations corroborate this choice, showing that one localized orbital at
each magnetic ion indeed is enough to describe the low energy physics of the system.
We will comment later on the possibility of considering a more
complex Hubbard model to describe the non-magnetic bridges between
Cu atoms, as well as the need of including more than one orbital at
the Cu sites.

The second quantized one-band Hubbard Hamiltonian reads
\begin{equation}
H_U = -t\sum_{i,j }\sum_{\scriptstyle  \alpha}
\Big\{c_{i\alpha}^{\dagger}c_{j\alpha}^{\phantom{\dagger}}+
{\rm h. c.}\Big\} +\frac{1}{2}U\sum_i n_{i\uparrow}\, n_{i\downarrow}\;,
\label{eq:Hu}
\end{equation}
where
$c_{i\alpha}^{\dagger}$
($c_{i\alpha}^{\dagger}$) creates (destroys) an electron with spin
$\alpha$ at  site $i$ and $n_{i\alpha}=c_{i\alpha}^{\dagger}c_{i\alpha}$  is the particle
number operator. More precisely the index $i$ labels a Wannier function localized at site $i$.
The first term represents the kinetic energy, characterized by a spin-independent ho\-pping parameter $t$,
which is the same for all pairs of site due to the $C_3$ symmetry of the Cu$_{3}$
molecule magnet.
The second-term is an on-site repulsion energy of strength $U$,
which has an effect only when two electrons of opposite spins reside
on the same site. It is the on-site repulsion energy.

The  spin-orbit interaction in the Hubbard model is described by
adding the following spin-dependent ho\-pping term~\cite{friedel_1964, kaplan_1982, bonesteel1992, Trif2010}
\begin{equation}
H_{\rm SOI}  = \sum_{i,j}\sum_{\scriptstyle  \alpha, \beta}
\Big\{ c_{i\alpha}^{\dagger}\Big(i\frac {{\bf P}_{ij}}{2}
\cdot {\pmb{ \sigma}}_{\alpha\beta}\;\Big) c_{j\beta}^{\phantom{\dagger}}+
{\rm h. c.}\Big\}\;,
\label{eq:hubbard_soi_gen}
\end{equation}
where $\pmb{\sigma}$ is the vector of the three Pauli matrices.
Here the vector ${\bf P}_{ij}$ is proportional to the matrix element
of $ {\pmb{ \nabla}} V\times {\bf p}$ between the orbital parts of
the Wannier functions at sites $i$ and $j$; $V$ is the one-electron
potential and $\bf p$ is the momentum operator. Clearly the
spin-orbit term has the form of a spin-dependent ho\-pping, which is
added to the usual spin-independent ho\-pping proportional to $t$.
This form of the spin-orbit interaction is a special case of
Moriya's ho\-pping terms\cite{Moriya1960} in the limit that all but
one orbital energy is taken to infinity.\cite{kaplan_1982}

In contrast to the spin-independent ho\-pping term, the spin-depending
ho\-pping parameters are related by both the full symmetry of the
molecule and the {\it local symmetry} of localized
orbitals.\cite{Trif2010} Now, because of the $\sigma_v$ symmetry,
${\bf P}_{ij} = P {\bf e}_z$. The final expression of the Hubbard
model, including the spin-orbit interaction is
%
%      Hubbard Hamiltonian with spin-orbit interaction
%
%
\begin{eqnarray}
H_{U+{\rm SOI}}  &=& \sum_{\scriptstyle i,\alpha}
\Big\{ c_{i\alpha}^{\dagger}\big (-t + i\lambda_{\rm SOI}\alpha\big )
c_{i+1\alpha}^{\phantom{\dagger}}+
{\rm h. c.}\Big\} \nonumber \\
&& + \frac{1}{2}U \sum_{\scriptstyle i}n_{i\uparrow}\, n_{i\downarrow}\;,
\label{eq:hubbard_soi}
\end{eqnarray}
where $\lambda_{\rm SOI} \equiv P/2 = {\bf P}_{ij}/ 2 \cdot {\bf e}_z$ is the
spin-orbit parameter.

We want to treat the two ho\-pping terms perturbatively on the same
footing, by doing an expansion around the atomic limit $t/U\;, \
\lambda_{\rm SOI}/U \to 0$. In many molecular magnets $t
\gg\lambda_{\rm SOI}$. This turns out to be the case also for
\{Cu$_3$\}. In other molecules the two ho\-pping parameters are of the
same order of magnitude.

We are interested in the half-filling regime. We
know that second-order perturbation theory in $t$ results in an
antiferromagnetic isotropic exchange term that splits the spin
degeneracy of the low-energy sector of the Hubbard model, defined by
the singly-occupied states. This action can be represented with an
effective spin Hamiltonian, the isotropic Heisenberg model, with
exchange constant $J = 4t^2/|U|$.\cite{Fradkin1991} Similarly Loss
{\it et al}. showed that another second-order term proportional to
$t\lambda_{\rm SOI}/U$ generates an anisotropic exchange term that
can be identified with the DM interaction.\cite{Trif2010} They write
approximate adapted many-body states to first-order in the
perturbation $|t|,\lambda_{\rm SOI} \ll U$, corresponding to
singly-occupied states. In particular there are two independent
doublets,
\begin{equation}
|\psi_{E'_{\pm}}^{1\, \alpha}\rangle =
\frac{1}{\sqrt{3}}\big (|\downarrow \uparrow \uparrow\rangle +
\epsilon_{\pm} |\uparrow\downarrow\uparrow\rangle + \epsilon_{\mp} |\uparrow\uparrow\downarrow\rangle\big )\;,
\label{eq:chiral_h_p}
\end{equation}
and
\begin{equation}
|\psi_{E'_{\pm}}^{1\, \alpha}\rangle=
\frac{1}{\sqrt{3}}\big  (|\uparrow\downarrow \downarrow \rangle +
\epsilon_{\pm} |\downarrow\uparrow\downarrow\rangle + \epsilon_{\mp} |\downarrow\downarrow\uparrow\rangle\big )
\;,
\label{eq:chiral_h_m}
\end{equation}
with $\epsilon_{\pm}=\exp\left( \pm 2\pi i/3  \right)$.
These are states with $S=1/2$ and $S_z  = \pm 1/2$.  These
states are formally identical to the  chiral states given in Eqs.
(\ref{eq:chi_pm_sz+}) and (\ref{eq:chi_pm_sz-}). Now, each of the
terms appearing in these equations is a single Slater determinant
obtained by three creation operators acting on the vacuum, e.g.,
\begin{equation}
|\uparrow \uparrow\downarrow\rangle \equiv
c_{1\,\uparrow}^{\dagger} c_{2\,\uparrow}^{\dagger} c_{3\,\downarrow}^{\dagger}
|0\rangle\;.
\end{equation}
The states $|\psi_{E'_+}^{1\, \alpha}\rangle$ and $|\psi_{E'_-}^{1\,
\alpha}\rangle$ are eigenstates of Hubbard Hamiltonian when $t=
\lambda_{\rm SOI} =0$. The tunneling and SOI mix the singly-occupied
and double-occupied states. The first-order correction is obtained
by mixing in doubly-occupied states
\begin{eqnarray}
|\Phi_{E'_{\pm}}^{1\, \alpha}\rangle
&\equiv &|\psi_{E'_{\pm}}^{1\, \alpha}\rangle
+ \frac {({\epsilon_-} -1)(t\pm \alpha \lambda_{\rm SOI})}{\sqrt{2}U}
 |\psi_{E^{'1}_{\pm}}^{2\, \alpha}\rangle \nonumber \\
 &&
+ \frac {3{\epsilon_+} (t\pm \alpha \lambda_{\rm SOI})}{\sqrt{2}U}
 |\psi_{E^{'2}_{\pm}}^{2\, \alpha}\rangle\;,
\label{eq:chiral_u1st}
\end{eqnarray}
where
\begin{equation}
|\psi_{E^{'1}_{\pm}}^{2\, \alpha}\rangle= \frac{1}{\sqrt{6}}\sum_{i=1}^3 \epsilon_{1,2}^{i-1}\left(
|\psi_{i1}^{\alpha}\rangle +|\psi_{i2}^{\alpha}\rangle
 \right)\;,
\end{equation}
and
\begin{equation}
|\psi_{E^{'2}_{\pm}}^{2\, \alpha}\rangle  = \frac{1}{\sqrt{6}}\sum_{i=1}^3 \epsilon_{1,2}^{i-1}\left(
|\psi_{i1}^{\alpha}\rangle -  |\psi_{i2}^{\alpha}\rangle
 \right)\;,
\end{equation}
with $|\psi_{ij}^{\alpha}\rangle = c_{i\uparrow}^{\dagger}
c_{i\downarrow}^{\dagger} c_{j\alpha}^{\dagger}|0\rangle$ ($i=1,2,3$
and $j\neq i$) representing the double-occupied sites.

The next step is to take the expectation value of the spin-orbit
part of Eq.(\ref{eq:hubbard_soi}) in these approximated states. The
result is\cite{Trif2010}
\begin{equation}
\langle\Phi_{E'_{\pm}}^{1\,\alpha}|H_{\rm SOI}|\Phi_{E'_{\pm}}^{1\,\alpha}\rangle
= \pm \frac {5 \sqrt{3} \lambda_{\rm SOI} t} {2U} {\rm sgn} (\alpha)\;.
\end{equation}

Note that off-diagonal matrix elements of $H_{\rm SOI}$ vanish; in
other words, SOI splits but does not mix the chiral states.

In the small $t/U \;, \ \lambda_{\rm SOI}/U $ limit, we can resort
to a spin-only description of the low-energy physics of the system.
The ground-state manifold (corresponding to the states
Eq.~(\ref{eq:chiral_u1st})) is given by the two chiral spin states
Eqs.~(\ref{eq:chi_pm_sz+}),~(\ref{eq:chi_pm_sz-}).

The anisotropic DM spin exchange Hamiltonian in $D_{3\text{h}}$
symmetry is given by~\cite{Trif2010}
\begin{equation}
H_{\rm DM}=  \frac{iD_z}{2}\sum_1^3(s^i_+\; s^{i+1}_- - s^i_-\;
s^{i+1}_+)
\label{eq:dm_spin}
\end{equation}

Now, in the low energy regime corresponding to a $D_{3 \text h}$
symmetric molecule magnet, the spin-orbit interaction can be reduced
to the  effective form
\begin{eqnarray}
H_{SOI}&=&  \Delta_{SOI} C_z S_z~\;,
\label{eq:Hsoitriangle}
\end{eqnarray}
where $\Delta_{SOI}$ is the effective SOI coupling constant.

The DM interaction expressed in this form clearly shows that it
splits but does not mix the two chiral states.\footnote{Note also
that the Hamiltonian in Eq.~\ref{eq:dm_spin} does not mix the GS
spin $S=1/2$ manifold with the excited-state spin $S= 3/2$ multiplet
and leaves the latter unchanged.} The splitting is exactly
proportional to $D_z$ and allows us, in the low-energy regime, to
make the identification
\begin{equation}
D_z  = \frac {5 \lambda_{\rm SOI} t} {U}~.
\label{eq:Dz}
\end{equation}

This Hubbard model analysis suggests an avenue to extract the DM
parameters from an ab-initio calculation. Only three parameters are
needed, namely the spin-orbit interaction $\lambda_{SOI}$, the
ho\-pping parameter $t$ and the on-site repulsion energy $U$.

%%%%%%%%%%%%%%%%%%%%%%%%%%%%%%%%%%%%%%%%%%%%%%%%%%%%%%%%%%%%%
%                                                           %
%       subsection  Semiclassical analysis of the DM interaction in frustrated spin systems
%                                                           %
%%%%%%%%%%%%%%%%%%%%%%%%%%%%%%%%%%%%%%%%%%%%%%%%%%%%%%%%%%%%%
\subsection{Semiclassical analysis of the DM interaction in frustrated spin systems}
\label{sec:SemiclassicalanalysisoftheDMinfrustratedspinsystems}
The quantum mechanical frustration present in an antiferromagnetic
spin triangle and the DM interaction both tend to favor
non-collinear spin configurations. It is instructive to study their
interplay in a semiclassical approach, where non-collinearity is a
more intuitive concept.

\begin{figure}[h]
\subfigure[Left-handed]{\label{fig:Lefthanded}{\resizebox{1.5in}{1.5in}{\includegraphics{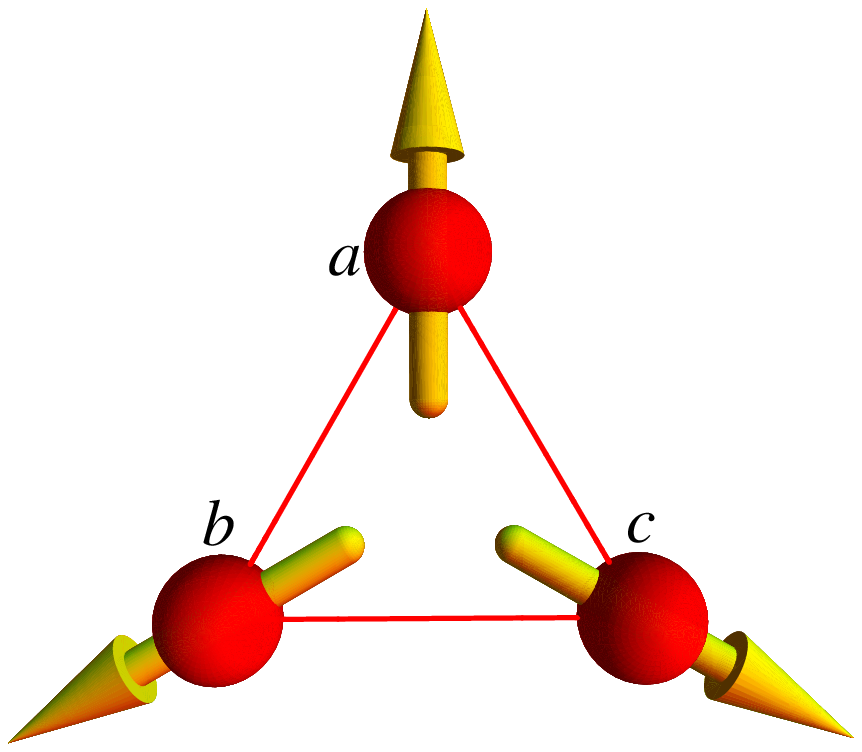}}}}
\subfigure[Right-handed]{\label{fig:Righthanded}{\resizebox{1.5in}{1.5in}{\includegraphics{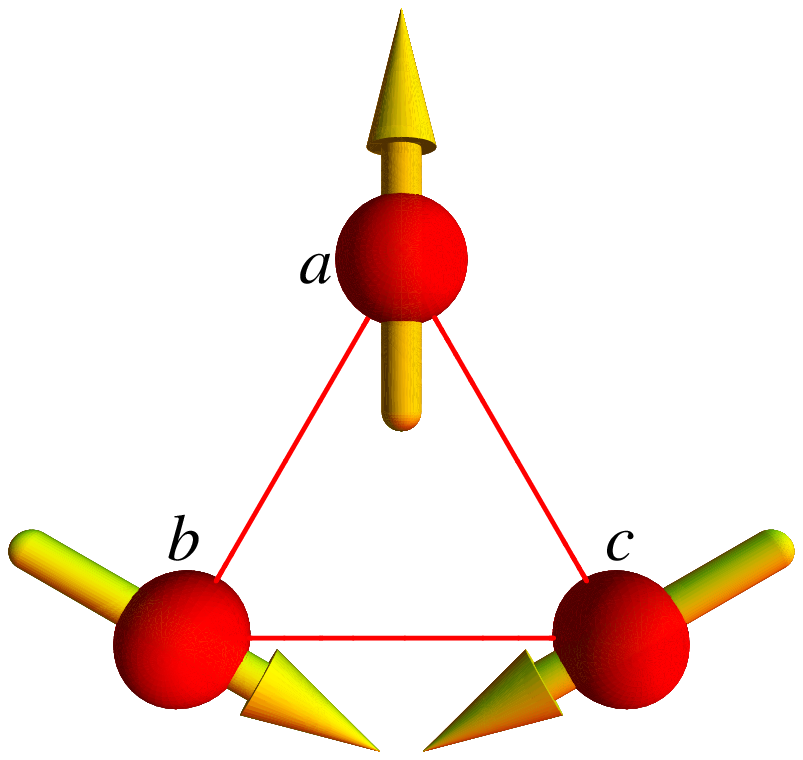}}}}
\caption{Non-collinear helical system of three Cu-atom spins. Anticlockwise rotations (by $- 240^0$)
define a left-handed helical state (a); clockwise
rotations (by $+ 240^0$) define a right-hand helical state (b).} \label{fig:triangleSpin}
\end{figure}

The {\it classical} Heisenberg model with energy functional given by
Eq.~(\ref{eq:afmh}) has two degenerate ``ground states'', given by
the two non-collinear spin configurations shown in
Fig.~\ref{fig:triangleSpin}.
Classically these two states are the best way to by-pass the
frustration present for any collinear spin configuration in a
triangular antiferromagnet. Quantum mechanically the two
non-collinear spin configurations can be re\-presented by the states
\begin{eqnarray}
|\psi_{{\rm nc}\, \pm} \rangle &=& \Big[ (\alpha_1
|\uparrow\rangle_1 + \beta_1|\downarrow\rangle_1 )\otimes
 (\alpha_2 |\uparrow\rangle_2  + \beta_2|\downarrow\rangle_2 )   \nonumber \\
 && \otimes
 (\alpha_3 |\uparrow\rangle_3 + \beta_3|\downarrow\rangle_3 ) \Big]\;,
\label{eq:ncs}
\end{eqnarray}
where $\alpha= \cos(\theta/2)$ and $\beta=\exp\{ i\phi
\}\sin(\theta/2)$. Here $\theta$ is the elevation angle and $\phi$
is the azimuth angle. The three spinors $(\alpha_i
|\uparrow\rangle_i + \beta_i|\downarrow\rangle_i )\;, \ i = 1, 2,3$
are three spin-$1/2$ coherent states defined by the three
non-collinear directions obtained by rotating consecutively by the
angle $\pm 240^0$ (see Fig. \ref{fig:triangleSpin}). Anticlockwise rotations (by $- 240^0$)
define a left-handed helical state (Fig. \ref{fig:Lefthanded}); clockwise
rotations (by $+ 240^0$) define a right-hand helical state (Fig.
\ref{fig:Righthanded}).
%
%
%                Triangle Figure
%
%

In contrast to the true GS given in Eqs.~(\ref{eq:chi_pm_sz+}),~(\ref{eq:chi_pm_sz-})
the non-collinear states defined in Eq.~(\ref{eq:ncs})
are neither eigenstates of the quantum Hamiltonian
Eq.~(\ref{eq:afmh}) nor of ${\bf S}^2$ and $S_z$. The expectation
value of the Hamiltonian $H_H$ at these states is defined by
\begin{equation}
\langle\psi_{{\rm nc}\, \pm}  | H_{\rm H}| \psi _{{\rm nc}\, \pm}
\rangle= 3J/4\;. \label{eq:e_ncs}
\end{equation}

The fact that the energy of the collinear states is higher than the
energy of the chiral states by $3J/8$ is not surprising, since the
noncollinear states defined in Eq.~(\ref{eq:ncs}) are a mixture of
$S=1/2$ and $S=3/2$ components.

When rewritten in term of the electronic states for the
corresponding Hubbard model at half-filling in the small $t/U$
limit, the non-collinear spin-coherent states defined in
Eq.~(\ref{eq:ncs}) can be considered to be the ``best'' energy states
given by a single Slater determinant (Note that the chiral states
cannot be written as a single Slater determinant).

It is now interesting to examine the effect of the DM interaction on
these states. A straightforward calculations shows that for the DM
interaction of Eq.~(\ref{eq:dm_spin}), where only the $z$-component
of ${\bf D}$ is nonzero
\begin{equation}
\langle\psi_{{\rm nc}\, \pm}  | H_{\rm DM}| \psi _{{\rm nc}\, \pm}
\rangle = \pm \frac{3}{4} \frac{\sqrt{3}}{2} D_z\;.
\label{eq:ev_dm_ncs}
\end{equation}

Therefore, as for the GS manifold of the exact eigenstates, the DM
interaction {\it splits} but does {\it not couple} the two
noncollinear states.
The DM para\-meter $D_z$ is, by Eq.~(\ref{eq:ev_dm_ncs}), related to
the DM interaction-induced energy-gap between the two noncollinear states
\begin{eqnarray}
\Delta E_{\rm nc} &=& \langle\psi_{{\rm nc}\, +}  | H_{\rm DM}| \psi
_{{\rm nc}\, +} \rangle -
\langle\psi_{{\rm nc}\, -}  | H_{\rm DM}| \psi _{{\rm nc}\, -} \rangle \nonumber \\
&=& \frac{3\sqrt{3}}{4} D_z\;.
\label{eq:ncs_gap}
\end{eqnarray}

This result suggests a way of extracting the DM vector parameter
${\bf D}$ similar in spirit to the method used to calculate the
isotropic exchange parameter $J$ by comparing the energy difference
of states with ferromagnetic and antiferromagnetic spin
configurations respectively. In the next section we will see that
this procedure can also be carried out by first-principle methods.

%%%%%%%%%%%%%%%%%%%%%%%%%%%%%%%%%%%%%%%%%%%%%%%%%%%%%%%%%%%%%
%                                                           %
%       section   ab initio calculation of the DM vector
%                                                           %
%%%%%%%%%%%%%%%%%%%%%%%%%%%%%%%%%%%%%%%%%%%%%%%%%%%%%%%%%%%%%
\section{Ab-initio calculation of the DM vector}
\label{sec:AbinitiocalculationoftheDMvector}

All the calculations in this work are carried out by using {\it ab
initio} package NRLMOL \cite{Pederson1990, Jackson1990}, which uses a
Gaussian basis set to solve the Kohn-Sham equations within PBE-GGA
approximation.\cite{Perdew} For more computational details and the
electronic properties of \{Cu$_3$\} we refer the reader to our
previous work.\cite{islam2010}

%%%%%%%%%%%%%%%%%%%%%%%%%%%%%%%%%%%%%%%%%%%%%%%%%%%%%%%%%%%%%
%                                                           %
%       subsection   t calculation
%                                                           %
%%%%%%%%%%%%%%%%%%%%%%%%%%%%%%%%%%%%%%%%%%%%%%%%%%%%%%%%%%%%%
\subsection{Calculation of the ho\-pping term $t$}
\label{sec:tCalculation}

As discussed in the section \ref{methods:3}, the Hubbard model approach is
based on allowing the localized electrons to hop to its nearest
neighbor sites and in the present case of the \{Cu$_3$\} molecule, these
localized electrons are $d$ electrons. Therefore, for calculating
ho\-pping parameter $t$, the relevant states are those $d$ electron
states that lie close to the Fermi level. Let $|K,\alpha\rangle$ be
the three relevant Kohn-Sham eigenstates calculated from NRLMOL. We
can write them as a linear combination of the localized atomic
orbitals, centered at the three Cu sites, $\{ \left| \phi_a
\right\rangle, \left| \phi_b \right\rangle , \left| \phi_c
\right\rangle\}\otimes \left| \chi_{\alpha} \right\rangle $, with
$\alpha=\uparrow,\downarrow$ for spin up and down, respectively:

\begin{equation}
\left|K,\alpha \right\rangle = \sum_{i} C^i_{K\alpha} \left| \phi_i  \right\rangle \left| \chi_{\alpha}  \right\rangle\;.
\label{eq:lc}
\end{equation}
where $C^i_{K\alpha}$ is the weight of the localized $\left| \phi_i  \right\rangle \left| \chi_{\alpha}  \right\rangle$ wavefunction.

\begin{figure}[h]
{\resizebox{2.0in}{1.0in}{\includegraphics{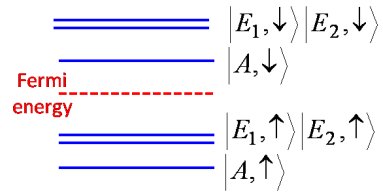}}}
\caption{Schematic diagram of the Kohn-Sham energy levels around the Fermi level}
    \label{fig:levels}
\end{figure}

For the $\left| \uparrow \uparrow \uparrow \right\rangle$ spin
configuration, in the absence of spin-orbit interaction, the
relevant three levels around the Fermi level are doubly and singly
degenerate. These levels are sketched in Fig. \ref{fig:levels}
%%%%%%%%%%%%%%%%%%%%%%%%%%%%%%%%%%%%%%%%%%%%%%%%%%%%%%%%
%
%                    FIGURE   Energy levels
%
%%%%%%%%%%%%%%%%%%%%%%%%%%%%%%%%%%%%%%%%%%%%%%%%%%%%%%%%5

We obtain the level structure by diagonalizing the three-site
Hamiltonian in the absence of the SOI:
\begin{equation}
H_0 = \varepsilon_0 \sum_{i}   \left| \phi_i \right\rangle   \left\langle \phi_i \right| -t\sum_{i \neq j }  \left| \phi_i \right\rangle  \left\langle \phi_j \right|\;,
\label{eq:H0}
\end{equation}
where $\varepsilon_0$ is the onsite energy, $t$ is the ho\-pping term
and $i,j=a,b,c$ represent the copper sites. We get the eigenvalues
$\varepsilon_0+t$ and $\varepsilon_0-2t$ for the two-fold and one-fold
degenerate  states, respectively. The Kohn-Sham eigenvectors can be
defined as a linear combination of the localized wavefunctions,
\begin{eqnarray}
\left|E_1,\uparrow \right\rangle &=& \frac{1}{\sqrt{2}}\left( \left| \phi_a \right\rangle -\left| \phi_b \right\rangle  \right)\left|\uparrow \right\rangle \;,\nonumber \\
\left|E_2,\uparrow \right\rangle &=& \frac{1}{\sqrt{6}}\left( \left|
\phi_a \right\rangle +\left| \phi_b \right\rangle -2 \left| \phi_c
\right\rangle  \right)\left|\uparrow\right\rangle \;,
\label{eq:evectors} \\
\left|A,\uparrow \right\rangle &=& \frac{1}{\sqrt{3}}\left( \left|
\phi_a \right\rangle +\left| \phi_b \right\rangle +\left| \phi_c
\right\rangle \right)\left|\uparrow \right\rangle\;. \nonumber
\end{eqnarray}

Now the localized states can be written in term of the Kohn-Sham functions
\begin{eqnarray}
\left|\phi_a \right\rangle \left|\uparrow \right\rangle&=&
 \frac{\left| A,  \uparrow \right\rangle}{\sqrt{3}}
+\frac{\left| E_1,\uparrow \right\rangle}{\sqrt{2}}
+\frac{\left| E_2,\uparrow \right\rangle}{\sqrt{6}}\;, \nonumber \\
\left|\phi_b \right\rangle \left|\uparrow \right\rangle&=&
 \frac{\left| A,  \uparrow \right\rangle}{\sqrt{3}}
-\frac{\left| E_1,\uparrow \right\rangle}{\sqrt{2}}
+\frac{\left| E_2,\uparrow \right\rangle}{\sqrt{6}}\;,
\label{eq:linearKohnSham} \\
\left|\phi_c \right\rangle \left|\uparrow \right\rangle&=&
  \frac{\left| A,  \uparrow \right\rangle}{\sqrt{3}}
-2\frac{\left| E_2,\uparrow \right\rangle}{\sqrt{6}}\;. \nonumber
\end{eqnarray}

Our calculations showed that these states are primarily localized on
the Cu atoms and have $d$ character. We have obtained the Kohn-Sham
eigenenergies for the one-fold and two-fold degenerate states
\begin{eqnarray}
\left\langle E_1,\uparrow \right|H_0   \left|E_1,\uparrow \right\rangle &=&
\frac{1}{2}\left( \left\langle \phi_a \right| -\left\langle \phi_b \right| \right) H_0
\left( \left| \phi_a \right\rangle -\left| \phi_b \right\rangle  \right) \nonumber \\
&=&\varepsilon_0+t\;,  \nonumber \\
\left\langle A,\uparrow \right| H_0  \left|A,\uparrow \right\rangle &=&
\frac{1}{3}\left( \left\langle \phi_a \right| +\left\langle \phi_b \right| +\left\langle \phi_c \right| \right)H_0
\nonumber \\
&&\left( \left| \phi_a \right\rangle +\left| \phi_b \right\rangle +\left| \phi_c \right\rangle \right)
\nonumber \\
&=&\varepsilon_0-2t \;. \label{eq:system}
\end{eqnarray}

From Eqs.~(\ref{eq:system}) we can finally evaluate the value  of the parameter $t$ as:
\begin{eqnarray}
t&=&\frac{1}{3}\left( \left\langle E_1,\uparrow \right|H_0 \left|E_1,\uparrow \right\rangle-\left\langle A,\uparrow \right| H_0  \left|A,\uparrow \right\rangle \right) \nonumber  \\
&=&50.84 \text{meV} \;. \label{eq:t}
\end{eqnarray}

%%%%%%%%%%%%%%%%%%%%%%%%%%%%%%%%%%%%%%%%%%%%%%%%%%%%%%%%%%%%%
%                                                           %
%       subsection   $\lambda_{SOI}$ calculation
%                                                           %
%%%%%%%%%%%%%%%%%%%%%%%%%%%%%%%%%%%%%%%%%%%%%%%%%%%%%%%%%%%%%
\subsection{Calculation of the spin-orbit interaction parameter $\lambda_{SOI}$}
\label{sec:lambdaSOCalculation}

Standard spin-orbit interaction representation for spherical systems
is given by
\begin{equation}
U_{so}(r,{\textbf L},{\textbf S}) = \frac{1}{2c^2} {\textbf S} \cdot {\textbf L} \frac{1}{r} \frac{d{\bf \Phi} (r)}{dr}\;,
\label{eq:cso}
\end{equation}
where $r$ is the position, ${\textbf L}$ is the angular momentum,
${\textbf S}$ is the spin moment, $c$ is the speed of light, and
$\Phi$ is a spherically symmetric potential. The above
equation is exact for spherical systems. For a multicenter system a superposition of such terms needs to be considered.
However, this
approximation could miss non-spherical correlations important for
anisotropic energies. Instead of using Eq.~(\ref{eq:cso}), a
generalization of the spin-orbit interaction for non-spherical or
multicenter systems is given by
\begin{equation}
U_{so}({\textbf r},{\textbf p},{\textbf S}) = - \frac{1}{2c^2} {\textbf S} \cdot {\textbf p} \times {\bf \nabla\Phi({\textbf r})}\;,
\label{eq:qso}
\end{equation}
where $\textbf p$ is the momentum operator and a external electric
field is given by ${\textbf E}=-{\bf \nabla\Phi}$.

Pederson {\it et. al} (see Ref. \onlinecite{Pederson1999}) have
%Pederson {\it et. al} (see Ref. \onlinecite{Pederson1999a}) have
shown an exact simplified method for incorporating spin-orbit
coupling into density-functional calculations. In order to get the
basis-set for the spin-orbit coupling the single-electron wave
function can be expressed as
\begin{equation}
\psi_{is}({\textbf r}) = \sum_{j\alpha} C_{j\alpha}^{is} f_{j}({\textbf r}) \chi_{\alpha}\;,
\label{eq:basisset}
\end{equation}
where $ f_{j}({\textbf r})$ is a spatial basis function,
$\chi_{\alpha}$ is either a majority or minority spin spinor, and $
C_{j\alpha}^{is}$ are determined by effectively  diagonalizing  the
Hamiltonian matrix. In order to calculate the effect of the SOI (Eq.
(\ref{eq:qso})) it is necessary to calculate matrix elements of the
form
\begin{eqnarray}
U_{j\alpha,k\alpha'} &=& \left\langle f_j \chi_{\alpha} \right| U({\textbf r},{\textbf p},{\textbf S})  \left|  f_k \chi_{\alpha'}  \right\rangle
\nonumber  \\
&=& \sum_x \frac{1}{i}
\left\langle f_j  \right| V_x  \left|  f_k  \right\rangle
\left\langle  \chi_{\alpha} \right| S_x \left|  \chi_{\alpha'}  \right\rangle \;,
\label{eq:matrixelements}
\end{eqnarray}
where
\begin{equation}
\left\langle f_j  \right| V_x  \left|  f_k  \right\rangle  = \frac{1}{2c^2} \left(
\left\langle    \frac{df_j}{dz}   \right|  { \mathbf \Phi}  \left|   \frac{df_k}{dy}      \right\rangle-
\left\langle    \frac{df_j}{dy}   \right|  { \mathbf \Phi}  \left|   \frac{df_k}{dz}      \right\rangle
\right) \;.
\label{eq:Vx}
\end{equation}

The matrix elements for $V_y$ and $V_z$ are obtained by cyclical
permutations of $x,y $ and $z$ in Eq.~(\ref{eq:Vx}). This
methodology for the SOI matrix gives  several advantages, namely, it
does not require the determination of the electric field; it is
specially ideal for basis functions constructed  from
Gaussian-type orbitals, Slater-type functions, and plane waves.

We are interested in the matrix elements in the localized basis-set,
Eq.~(\ref{eq:linearKohnSham}):
\begin{eqnarray}
\left\langle \phi_i \right| \left\langle \chi_{\uparrow} \right| U_{so}  \left|  \phi_k  \right\rangle  \left|  \chi_{\uparrow} \right\rangle &=&  - \frac{1}{2c^2}
\left\langle \phi_i \right|  {\textbf p} \times {\bf \nabla\Phi({\textbf r})} \left|  \phi_k  \right\rangle
\nonumber \\
&&\cdot \left\langle \chi_{\uparrow} \right|  {\textbf S} \left|  \chi_{\uparrow}  \right\rangle \nonumber \\
&=&   \frac{1}{2i}
\left\langle \phi_i \right|  V_z \left|  \phi_k  \right\rangle \nonumber \\
&=&  - \frac{i}{2}   p_{ik}^z \equiv -i \lambda_{SOI}\;.
\label{eq:solbs}
\end{eqnarray}

We can write these matrix elements in the Kohn-Sham basis set
\begin{eqnarray}
\left\langle \phi_i \right| \left\langle \chi_{\uparrow} \right| U_{so}  \left|  \phi_k  \right\rangle  \left|  \chi_{\uparrow} \right\rangle &=&
\sum_{KK'}\left( \tilde{C}_{K\uparrow}^i \right)^* \tilde{C}_{K'\uparrow}^i \nonumber \\
&&\times \left\langle K,\uparrow \right|  U_{SOI} \left| K',\uparrow
\right\rangle ~. \label{eq:sokhbs}
\end{eqnarray}

We have obtained the matrix elements for the spin-orbit
interaction in the Kohn-Sham basis, $\{ \left| E_1 \right\rangle ,
\left|E_2 \right\rangle ,    \left| A \right\rangle \}\otimes
\left| \chi_{\alpha}  \right\rangle$ (Eq.~(\ref{eq:sokhbs})), and
used Eqs.~(\ref{eq:linearKohnSham}) to obtain the matrix elements:
\begin{equation}
p^z=
\begin{pmatrix}
0&0.85&0.85\\
0.85&0&0.85\\
0.85&0.85&0\\
\end{pmatrix}\;.
\end{equation}

From Eq.~(\ref{eq:solbs}) we have $\lambda_{SOI}=p_{ik}^z/2=0.43$ meV.

%%%%%%%%%%%%%%%%%%%%%%%%%%%%%%%%%%%%%%%%%%%%%%%%%%%%%%%%%%%%%
%                                                           %
%       subsection   $U$ and $D_z$ calculation
%                                                           %
%%%%%%%%%%%%%%%%%%%%%%%%%%%%%%%%%%%%%%%%%%%%%%%%%%%%%%%%%%%%%
\subsection{Calculation of the Hubbard $U$ and evaluation of $D_z$ and $J$}
\label{U-Dz-Calc}

The most common approach for calculating $U$ involves calculation of
energy, $E$, of the molecule with $N$, $N+1$ and $N-1$ electron and extracting
U from the equation below,
\begin{eqnarray}
U&=&E(N+1)+E(N-1)-2E(N) \nonumber \\
 &=&[E(N+1)-E(N)]-[E(N)-E(N-1)] \nonumber \\
 &=& A-I \;.
\label{eq:Uconventional}
\end{eqnarray}

In the above equation A is (minus) the electron
affinity\footnote{Note that usually, the electron
affinity is defined as $[E(N)-E(N+1)]$, where $E(N)]$ is the energy of the neutral system.}
and I is the
ionization energy. For systems that are not closed shell, such as
those considered here, the $U$ value is essentially the second
derivative of energy with respect to charge and it is possible to
determine $U$ by calculating the energy as a function of charge.

For the single-band Hubbard-model corresponding to the \{Cu$_3$\}
molecule, we are interested in obtaining energies for the
charge-transfer excitations involving the transfer of a localized
{\it d}-electron on one copper site to a localized {\it d}-electron
on another site. Specifically we wish to know the energy of $\left|
X \right\rangle
=\left|\uparrow_a\downarrow_a\uparrow_c\right\rangle$ relative to
$\left| \uparrow_a\downarrow_b\uparrow_c \right\rangle $. There are
a total of twelve charge-transfer excitations that can be made with
one-site doubly occupied and one electron on one of the other sites.
For the half-filled case of interest here, the energy difference
depends upon the electron affinity of the state on site $a$, the
ionization energy of the state on site $b$ and the residual
long-range coulomb interaction between the negatively charged
electron added to site $a$ and the positively charged hole that is
left behind on site $b$. Since site $b$ and site $a$ are equivalent,
it follows that we simply need to calculate $U$ for any one of the
copper sites in the half filled case. A very rough estimate of the
charge transfer energy may be determined by calculating the PBE-GGA
energy of the Cu atom with an electron configuration of
$1s^22s^23s^24s^2p^63p^63d^n$ with $n$=8,9,10. Using $n$=9 as the
reference state, one finds a bare $U$ value of 13.76 eV which, after
accounting for the particle-hole interaction
($27.2116/R_{Cu-Cu}=2.95 eV$, where $R_{Cu-Cu}=4.87$ Bohr is the
distance between magnetic centers) is shifted to 10.8 eV.

\begin{figure}[h]
{\resizebox{3.2in}{2.2in}{\includegraphics{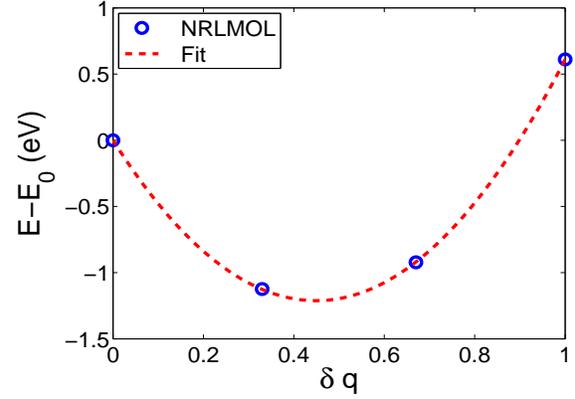}}}
\caption{(Color online) Dependence of the total energy on added fractional
charge $\delta q$. The (blue) circle represent the results of NRLMOL calculations and the
dashed (red) line represents a quadratic fit.}
\label{fig:Evsq}
\end{figure}

In the \{Cu$_3$\} molecule, we have chosen to calculate $U$
quasi-analytically by gradually adding (or subtracting) a small
fraction of electronic charge $\delta q$ to one of the half-filled
Cu $d$-states. The energy of the system as a function of $\delta q$
is shown in Fig. \ref{fig:Evsq}, where we can see that it can well be reproduced by a
quadratic fitting curve.
The figure shows that, upon adding a fractional charge to a localized orbital, the
total energy initially decreases, since the orbital energy is negative. Eventually, however, the
competing Coulomb repulsion takes over and the net change in total energy for adding one electron
to a localized orbital is positive. In contrast, with one extra electron delocalized throughout the molecule
the total energy is usually smaller than the energy of the neutral molecule.
%%%%%%%%%%%%%%%%%%%%%%%%%%%%%%%%%%%%%%%%%%%%%%
%
%     Figure   U calculation
%
%%%%%%%%%%%%%%%%%%%%%%%%%%%%%%%%%%%%%%%%%%%%5

The difference in the energy of the system before and after adding a
fraction of electronic charge $\delta q$ is given by $\Delta E =
U_{eff} = U\delta q^2 - e^2 \delta q^2 / R_{\text{Cu-Cu}}$, where
$U=\partial^2 E(q)/\partial q^2$. We have calculated the effective
parameter $U_{eff}$ by setting $\delta q = 1$:
\begin{eqnarray}
U_{eff}&=& \delta q^2 \left( \frac{\partial^2 E(q)}{\partial q^2}-\frac{e^2}{R_{{\text{Cu-Cu}}}} \right) \nonumber \\
&=& 9.06  {\text{ eV}} \;, \label{eq:ourU}
\end{eqnarray}
where $E(q)=E_0+(U/2)(q-q_0)^2$ with $E_0$ being a constant.

\subsubsection{Evaluation of $D_z$ and $J$}
Having calculated the parameters $t$, $\lambda_{SO}$, and
$U_{eff}$, we are now able to use Eq.~(\ref{eq:Dz}) and  evaluate the
Dzyaloshinskii-Moriya parameter $D_z$. We obtain

\begin{equation}
D_z= 5 \frac{\lambda_{\rm SOI} t} {U_{\rm eff}}=0.01 {\rm meV}\;.
\end{equation}

This value of $D_z$ yields a small splitting of the chiral state, $\Delta \approx 0.02$ meV
$\approx 0.3$ K. Experimental estimates of the DM parameter %CMC: Need a reference here
find a splitting 3-4 times larger than this value. Considering the
smallness of this energy and the uncertainty in the experimental
measurements, the two estimates are consistent with each other. On
the other hand, it is also possible that part of the discrepancy
between theory and experiment is due to the fact that other
mechanisms, different from the DM interaction, contribute to the
splitting. In particular Ref.\cite{furukawa_PRB_2007} pointed out
that small deformations of the triangular molecule can lift the
chiral degeneracy and this contribution to the splitting  could be
even more important than the DM interaction. If this is indeed the
case, our results would imply that our method of computing the DM
parameter is actually rather accurate.

From a computational point of view, it is interesting at this point
to evaluate the isotropic exchange constant $J$ from the Hubbard
model perturbative approach, which gives

\begin{equation}
J = 4t^2/U \approx  1 {\rm meV}.
\end{equation}

This estimate of $J$ is considerably closer to the experimental
value of 0.5 meV than the value of 3.7 meV obtained by computing the
energy difference between states with ferromagnetic and
antiferromagnetic spin configurations.\cite{islam2010}

%%%%%%%%%%%%%%%%%%%%%%%%%%%%%%%%%%%%%%%%%%%%%%%%%%%%%%%%%%%%%
%                                                           %
%       subsection             Comparison with other methods
%                                                           %
%%%%%%%%%%%%%%%%%%%%%%%%%%%%%%%%%%%%%%%%%%%%%%%%%%%%%%%%%%%%%
\subsection{Comparison with other methods}
\label{sec:othermethods}

In a recent work Takeda {\it et. al.}\cite{takeda2007} have used a
non-collinear approach to estimate the DM interaction.  Instead of
the use of simple product functions, this work capitalizes on the
use of generalized orbitals which are composed of a linear
combination of both spinors with different and variable spatial
functions.  By using such a representation it is possible to develop
single-determinants which are composed of a linear combination of
the chiral spin 1/2 states and the non-chiral spin 3/2 states. For
example, the states associated with the system depicted in Fig
\ref{fig:Lefthanded} would be represented according to:\\
\begin{eqnarray}
\left| \psi_{nc\pm} \right\rangle& = & \left| X_{\pm}^aX_{\pm}^bX_{\pm}^c  \right>  \nonumber \\
 & = & \frac{1}{2\sqrt{2}} \left(
  \left|\uparrow\uparrow\uparrow\right>
\pm i\left|\downarrow\uparrow\uparrow\right>
\mp(-1)^{1/6}\left|\uparrow\downarrow\uparrow\right>       \right. \nonumber \\
&&\left. \mp(-1)^{5/6}\left|\uparrow\uparrow\downarrow\right>
\mp i\left|\downarrow\downarrow\downarrow\right>
-\left|\uparrow\downarrow\downarrow\right>      \right. \nonumber \\
&&\left.
+(-1)^{1/3}\left|\downarrow\uparrow\downarrow\right>
-(-1)^{2/3}\left|\downarrow\uparrow\downarrow\right>   \right) \label{eq:noncollstates}
\end{eqnarray}
where $X_{+}(\theta,\phi)=\cos(\theta/2)\left|\uparrow \right> +\exp\{i\phi\}\sin(\theta/2)\left|\downarrow \right>$ and
$X_{-}(\theta,\phi)=\sin(\theta/2)\left|\uparrow \right> -\exp\{i\phi\}\cos(\theta/2)\left|\downarrow \right>$, with $\theta=\pi/2$ and $\phi=\pi/2,7\pi/2,-\pi/2$.
They further claim that $\Delta E_{\rm nc} = 3\sqrt{3}/4 D_z$
(see Eq.~\ref{eq:ncs_gap}) can be
estimated by a perturbational treatment of the SOI, as follows
\begin{equation}
\Delta E_{\rm nc} = \langle\psi_{{\rm nc}\, +}  | H_{\rm SOI}| \psi
_{{\rm nc}\, +} \rangle - \langle\psi_{{\rm nc}\, -}  | H_{\rm SOI}|
\psi _{{\rm nc}\, -} \rangle\;,
\end{equation}
where $H_{\rm SOI}$ is the one-electron spin-orbit interaction.
These expectation values can be calculated by DFT.

It is clear from the expression Eq. (\ref{eq:noncollstates}) that
the expectation value of the spin-orbit interaction for this and
other states would be linear so, without other considerations, one
can not extract an interaction that depends upon the excitations of
interest to the Hubbard Hamiltonian.  However, in analogy to the
expansion of the many-electron wavefunction for molecular hydrogen
in regions intermediate between the bonding and separated-atom
limit, a self-consistent optimization of such a starting determinant
allows the spin-orbitals to be intermediate between the doubly
occupied and single occupied representations. While the resulting
noncollinear wavefunction is still a single Slater determinant in
character, expansion of the noncollinear state in terms of the
Hubbard states would show a wavefunction comprised primarily of the
8$\times$8 half-filled determinants but would also contain small
contributions of the ionic contributions  which are shifted upward
by $U_{eff}$. It is the small admixture of these states that allow
Takeda {\it et. al.} to extract both the exchange parameters and the DM
interaction through the use of noncollinear representations. This
approach could have advantages from an operational viewpoint since
it effectively addresses the potential role of other excited states
that are routinely  excluded from the Hubbard Hamiltonian. However,
the precise interactions which ultimately mediate the appearance of
the DM interaction require additional analysis which is every bit as
arduous as that presented here.

%%%%%%%%%%%%%%%%%%%%%%%%%%%%%%%%%%%%%%%%%%%%%%%%%%%%%%%%%%%%%
%                                                           %
%       subsection       Green's function approach
%                                                           %
%%%%%%%%%%%%%%%%%%%%%%%%%%%%%%%%%%%%%%%%%%%%%%%%%%%%%%%%%%%%%

An alternative method to calculate the DM vector, based on
Andersen's ``local force theorem" \cite{andersen80}, was developed
by Solovyev {\it et al}.\cite{solovyev96} More recently this method
was utilized in conjunction with DFT to study the DM interaction
between magnetic atoms inserted in different crystalline systems and
surfaces. \cite{anisimov05,lichtenstein09}
%CMCfinal
Essentially this method expresses the DM vector in terms of the
Green's functions of the system, modified by the spin-orbit
interaction. Although computationally sophisticated, the Green's
function method is physically less transparent than the one adopted
here, particularly for a finite system such a triangular SMM, where
the crucial ingredients leading to the anisotropic DM exchange can
be reduced to a few parameters that have a direct physical
interpretation within the Hubbard model.

%%%%%%%%%%%%%%%%%%%%%%%%%%%%%%%%%%%%%%%%%%%%%%%%%%%%%%%%%%%%%
%                                                           %
%       section   Conclusions
%                                                           %
%%%%%%%%%%%%%%%%%%%%%%%%%%%%%%%%%%%%%%%%%%%%%%%%%%%%%%%%%%%%%

\section{Conclusions}
\label{summary}

We carried out a first-principles investigation of the zero-field
splitting of the chiral ground states of a \{Cu$_3$\}
single-molecule magnet (SMM), caused by the Dzyaloshinskii-Moriya
interaction. Our approach relies on the perturbative analysis of a
Hubbard model, which includes spin-orbit interaction. In the large
$U$ limit, appropriate for \{Cu$_3$\}, it is possible to express the
Dzyaloshinskii-Moriya constant in terms of the parameters that
define the Hubbard model, such as the effective ho\-pping integral
between magnetic sites $t$, the on-site repulsion energy $U$, and
the strength of the spin-orbit $\lambda_{\rm SOI}$. We then carried
out an approximate method to extract the values of these parameters
from our spin density functional theory calculations of the SMM. The
value of the Dzyaloshinskii-Moriya constant $D$ that we found is of
the order of $0.01$ meV, which is a factor of 5 smaller than the
value measured experimentally. Given the uncertainty of the
experimental result and the fact that other effects might contribute
to the zero-field spin splitting of the chiral states, our estimate
should be considered consistent with experiment.

The method of computing the DM parameter by effectively extending
Anderson's theory of superexchange to include spin-orbit interaction
is very close to Moriya's original formulation of anisotropic
exchange. It is interesting to note that if we use this approach to
calculate the isotropic superexchange constant $J$ of the Heisenberg
model describing \{Cu$_3$\}, we obtain a value that is closer to
experimental result than the estimates based on total energy
calculations of ferromagnetic vs antiferromagnetic spin
configurations. This seems to suggest that this approach is not only
physically very intuitive, but it might also bear promise of good
numerical accuracy.

While the methods discussed here provide physical insight into the
nature of the DM interaction, we note that for future calculations it
would be desirable to consider excitations that are not normally
included in the single-band Hubbard model. For such an approach it
would be necessary to include methodologies that allow for the
calculation of all excitations in such systems.

%%%%%%%%%%%%%%%%%%%%%%%%%%%%%%%%%%%%%%%%%%%%%%%%%%%%%%%%%%%%%
%                                                           %
%   section      Acknowledgment
%                                                           %
%%%%%%%%%%%%%%%%%%%%%%%%%%%%%%%%%%%%%%%%%%%%%%%%%%%%%%%%%%%%%

\section*{Acknowledgment}
\label{sec:Acknowledgment} This work was supported by the School of
Computer Science, Physics and Mathematics at Linnaeus University,
the Swedish Research Council under Grants No: 621-2007-5019 and 621-2010-3761,
and the NordForsk research network 080134 ``Nanospintronics: theory and simulations".
We would like to thank D. Loss and D. Stepanenko for very a helpful explanation of their Hubbard model
approach to the DM interaction in molecular antiferromagnets.
The early parts of this collaboration were
supported in part by NRL.

%%%%%%%%%%%%%%%%%%%%%%%%%%%%%%%%%%%%%%%%%%%%%%%%%%%%%%%%%%%%%
%                                                           %
%                        bibliography
%                                                           %
%%%%%%%%%%%%%%%%%%%%%%%%%%%%%%%%%%%%%%%%%%%%%%%%%%%%%%%%%%%%%

\bibliography{SOIpaper}

\end{document}